\documentclass[aps,prd,twocolumn,showpacs,amsmath]{revtex4}
\usepackage[dvips]{color,graphicx}
\usepackage{amsfonts}
\usepackage{amssymb}
\usepackage{latexsym}

\newcommand{\dalm}{\kern1pt\vbox{\hrule height 0.9pt\hbox{\vrule width
0.9pt\hskip 2.5pt\vbox{\vskip 5.5pt}\hskip 3pt\vrule width 0.3pt}\hrule height
0.3pt}\kern1pt}

\newcommand{\lw}[1]{\smash{\lower2.ex\hbox{#1}}}

\begin{document}


\title{On the enigmatic $\Lambda$ - a true constant of spacetime}
\author{Naresh Dadhich\thanks{Electronic address:nkd@iucaa.ernet.in}}
\email{nkd@iucaa.ernet.in}
\affiliation{Inter-University Centre for Astronomy \& Astrophysics, Post Bag 4, Pune~411~007, India\\}
\date{\today}

\begin{abstract} 
Had Einstein followed the Bianchi differential identity for the derivation of his equation of motion for gravitation, $\Lambda$ would have emerged as a true new constant of spacetime on the same footing as the velocity of light? It is then conceivable that he could have perhaps made the most profound prediction that the Universe may suffer accelerated expansion some time in the future! Further we argue that its identification with the quantum vacuum energy is not valid as it should have to be accounted for like the gravitational field energy by enlarging the basic framework of spacetime and not through a stress tensor. The acceleration of the expansion of the Universe may indeed be measuring its value for the first time observationally.
\end{abstract}

\pacs{04.20.-q, 04.20.Cv, 98.80.Es, 98.80.Jk, 04.60-m} 

\maketitle


The so called cosmological constant, $\Lambda$, as we all know, had a very chequered history mainly because the way Einstein introduced it in an adhoc manner as an after thought for having a static model of the Universe. Once the non-static solution for the Universe was found, its presence was no longer required for the purpose for which it was introduced. The situation remained ambiguous and it was left to one's taste to include it or not until the quantum  vacuum fluctuations were considered which relative to flat spacetime have the same form for the stress tensor as $\Lambda$. That is how it gets slated against the Planck length and to the annoying mismatch of 120 orders of magnitude. It again rises to describe the observed acceleration of the expansion of the Universe \cite{accel}. Notwithstanding the very ingenious and large number of dark energy models as well as the models involving modification of gravity and inhomogeneity, the observational data agrees with $\Lambda$ perfectly well. It is against this background we would like to go back right at the very beginning where it emerges in the most natural manner in the geometric derivation of the Einstein equation. \\

Bianchi and Rovelli have recently done a good bit of plain speaking on this very emotive issue \cite{br}. Though I quite resonate with their viewpoint yet I would like to expound upon some important aspects which have so far remained rather unnoticed or not fully appreciated. It would bring upon to bear an illuminating and insightful new perspective. It appears to me that had Einstein followed the Bianchi indentity for derivation of his equation of motion for gravitation, the situation would have been very different. Then $\Lambda$ would have entered into the equation at the same footing as the energy momentum tensor $T_{ab}$ and would have enjoyed its rightful place and meaning as a new constant of spacetime structure like the velocity of light, $c$. It really represents the \textit{all matter/force-free} state of spacetime which would have constant curvature. This is the most natural geometry it would have while any other including flat would be an external imposition. We adhere to the fundamental principle that since spacetime is a universal entity, \textit{nothing should be imposed on it}. Dynamically this state is characterized by homogeneity and isotropy of space and homogeneity of time, which should mean spacetime curvature must also be homogeneous (constant) and $\Lambda$ is then its measure. Even without reference to gravity, it thus naturally emerges as a characteristic of spacetime which is free of all matter and dynamics. \\

We shall also argue that its identification with the vacuum energy is rather misplaced because it has no independent existence like gravitational field energy of its own and is instead created by matter as a secondary source. Hence it should not sit alongside the primary matter source and gravitate through a stress. There have also been some considerations involving tracefree gravitational equation \cite{wein, paddy, ellis} that argue that it does not gravitate. On the contrary we adhere to the basic property of gravity of its universal linkage, hence gravitate it must but rather differently like the gravitational field energy \cite{dad-sch} that gravitates by curving the $3$-space. It can therefore only gravitate through enlargement of the spacetime framework as we go from classical to quantum gravity. Thus for its correct and true gravitational interaction there seems to be no shortcut but to wait for quantum theory of gravity to show the enlarged framework.  \\

Gravity is a universal force which means it links to all that physically exist in any form including the zero mass particles. In particular its linkage to the latter  requires that it can only be described by the spacetime curvature \cite{dad}. That is, its dynamics now resides in the spacetime curvature, $R_{abcd}$, and that should entirely determine it. We have no freedom to prescribe a law of gravity - it should all follow from the curvature. And it does indeed do that by the Bianchi differential identity, $R_{ab[cd;e]} = 0$. Taking its trace leads to the divergence free Einstein tensor, ($G_a{}^b{}_{;b} = 0$), and consequently to the equation,  
\begin{equation}
G_{ab} = \kappa T_{ab} - \Lambda g_{ab}, ~~~ T^{a}{}_{b;a} = 0  
\end{equation} 
where $G_{ab} = R_{ab} - \frac{1}{2} Rg_{ab}$ and $\Lambda g_{ab}$ is a constant relative to the covariant derivative. This becomes the equation for gravitation when we identify $T_{ab}$ with the energy-momentum distribution - a universal property which is shared by all that physically exist (universal source for the universal force). Here $\Lambda$ enters the equation on the same footing as $T_{ab}$ and hence cannot simply be wiped out at one's whims and fancy without proper physical explanation and justification. Had Einstein followed this route, it won't have been then an addition as an after thought for obtaining the static cosmological solution of the equation but would have rather emerged as a new constant of the Einsteinian gravity and spacetime? Is this procedure general enough, what happens when we include higher powers of curvature? Yes, it is, I have recently shown that the trace of the Bianchi derivative of a homogeneous polynomial in Riemann tensor does indeed yield the corresponding divergence free tensor, analogue of the Einstein  tensor \cite{dad1}. This is in fact a characterization of the Lovelock polynomial action. \\

A field equation has one free parameter that is determined by experimentally measuring the strength of the force. That parameter here is $\kappa$ and it is determined in terms of the Newtonian constant as $8\pi G/c^2$. What does this additional constant $\Lambda$ signify and how does one measure it? The feature that distinguishes gravity from all other forces is the absence of fixed background spacetime to which gravitational dynamics could be referred to, instead the dynamics now resides in spacetime itself. That means the equation must also provide the measure of zero gravitational dynamics, absence of gravity and the new constant provides that. $\Lambda$ thus defines the zero relative to which gravity should be measured. Now if Einstein sought static solution that would have simply indicated that the solution existed only for the specific value of $\Lambda$. The discovery of non-static solution by Friedmann would have created no issue at all. It would have shaved us from a great deal of misunderstanding and controversy. Like any other constant in the equation of motion, $\Lambda$ has to be empirically determined only by observations. That is what accelerating expansion of the Universe seems to be doing \cite{accel}. \\

The spacetime has to curve for describing gravitational field and so the question arises what should be its structure so that it can bend like all material objects. At deep down matter has discrete micro-structure, should spacetime also share this basic property with matter? Perhaps it is essential for it to be able to curve. The canonical quantization of gravitation - loop quantum gravity is of course based on the discrete microstructure \cite{abhay}. The next question is left to itself what geometry (flat or curved) would such a structure have \cite{ride}? In completely free state it should be maximally symmetric which would mean it should be of constant curvature. That is how $\Lambda$ is related to the basic structure of spacetime and it is indeed the measure of its affinity to curve in completely free state. \\

Let us now turn to the vacuum energy. When it is evaluated by using the flat spacetime quantum field theory (QFT) it has stress tensor that is proportional to the metric and that is how it gets associated with $\Lambda$. Firstly this is a flat space calculation that may or may not hold good in curved spacetime. More importantly the question is whether it is a right thing to include it in the equation at all? This is similar to asking for inclusion of the gravitational field energy in the Einstein equation. This we all know is meaningless. The general principle is that stress tensor includes only the primary source, matter energy momentum distribution and it is that alone should be on the right side of the equation. The secondary effects created by the primary source like the gravitational field energy and similarly the vacuum energy should not be included in the equation alongside the primary source energy momentum represented by $T_{ab}$. We shall demonstrate below how the gravitational field gravitates in the Einstein gravity? Something similar would have to be sought for the vacuum energy. Unfortunately that would come about only when we have a quantum theory of gravity. \\


The Einstein gravity distinguishes itself by the property of self interaction. How  is it incorporated in general relativity (GR) is what we would now show \cite{dad-sch}. Let us seek the vacuum solution for a mass point by writing the spherically symmetric static metric,
\begin{equation}
ds^2 = B dt^2 - A dr^2 - r^2 d\Omega^2. 
\end{equation} 
Then we have 
\begin{equation}
R^0_0 = -\frac{1}{2AB}\big(\nabla^2B - \frac{B^\prime}{2}(\frac{A^\prime}{A} +\frac{B^\prime}{B}) \big),
\end{equation}
\begin{equation}
R^1_1 = R^0_0 + \frac{1}{Ar}(\frac{A^\prime}{A} +\frac{B^\prime}{B}). 
\end{equation}
Clearly $R^0_0 = R^1_1$ implies $AB = const. = 1$ for asymptotic flatness, then $R^0_0 = 0$ reduces to $\nabla^2B = 0$, the same good old Laplace equation. What has then happened to the self interaction involving square of the field strength? If we set $A = 1$ (space part being flat), then we would have for $R^0_0 = 0$ as $\nabla^2B = {B^\prime}^2/2B$, showing the contribution of gravitational field energy. Instead what happens in GR is that contribution of gravitational field energy is incorporated in curving the space through $A = 1/B \neq 1$ leaving the Laplace equation intact to give the same Newtonian potential. This is how the gravitational field energy gravitates in GR through curving $3$-space leaving the potential unaffected. It is the enlargement of framework (from $3$-space flat to curved) that accounts for gravitational interaction of the gravitational field energy. This is what should happen for all secondary sources and not by writing a stress tensor on the right of the equation. Since vaccum energy is also a secondary source, it can honestly gravitate only through enlargement of spacetime framework. That we would know, since it is a quantum phenomena, only when there emerges a quantum theory of spacetime/gravity. \\

Note that the fluid equation of motion is given by $(\rho+p) \dot u_a = \nabla_bp(g_a^b - u_au^b)$ which clearly indicates $\rho+p$ as the inertial density. The vacuum energy has the equation of state, $\rho+p = 0$ implying vanishing of the inertial density. Whenever an inertial entity vanishes, it signals inapplicability of the existing law of motion as was the case for zero mass particle for Newton's Second law. That led to new theories of relativity, first special relativity (SR) to incorporate its universally constant velocity and then GR for making it feel gravity \cite{dad}. Once again there is an inertial entity vanishing and that should therefore ask for a new theory. That is the vacuum energy can be addressed in right earnest only by a new theory and so we again reach the same inference that it is for quantum gravity to incorporate the vacuum energy. \\


Now since $\Lambda$ is completely free of the vacuum energy, it could have any value. As a homogeneous curvature of \textit{free spacetime}, it should roughly measure its physical extent, the radius of curvature. However its actual value would be determined by the matter content, the energy density of the universe through the Einstein equation. For generic estimate, we could use the Einstein universe which would give its critical value as $\Lambda = \Lambda_c = 4\pi G \rho_0/c^2$ where $\rho_0$ is the present value of the density. For the spherical universe ($k=1$), this critical value is the separater between ultimate disperssion (for $\Lambda > \Lambda_c$) resembling the de Sitter universe and the recollapsing Friedmann universe (for $\Lambda < \Lambda_c$) \cite{jvn}. Without $\Lambda$ it is the curvature parameter $k$ that serves as the separater but now $k=1$ does not guarantee recollapse unless $\Lambda < \Lambda_c$. That is, there can be asymptotically dispersive universe for $k=1$ with a quasistationary phase if $\Lambda > \Lambda_c$. Since $\Lambda > 0$ has dispersive effect on matter, it has therefore to be bounded from above so as to make the universe recollapse. There is however a remarkable coincidence that $\Lambda_c$ corresponding to the present value of the density explains perfectly well the observed accelerating expansion of the Universe \cite{accel}. This is the simplest explanation and there is no need for the exotic dark energy. However it is not clear to what profound message, if any,  this coincidence is pointing?\\

We have argued that had Einstein followed the Bianchi identity route for derivation of his gravitational equation, $\Lambda$ would have been recognized as a true constant of spacetime on the same footing as the velocity of light. These are the only two constants that have been synthesized in the spacetime structure and no other constant could claim this degree of fundamentalness. The velocity of light binds space and time into a spacetime manifold while $\Lambda$ curves it. It is the measure of innate \textit{elastic} property of spacetime to curve even in the absence of all matter and gravity. This is the basic property of spacetime having  discrete microstructure which would in general have non-zero constant curvature \cite{ride} in the free state characterized by absence of all forms of matter and energy. Then introduction of matter simply breaks the constancy of curvature. On the other hand if free spacetime is taken as flat, the introduction of matter would ask for radical change in its structure so that it is able to curve. However dynamically free state is defined by maximal symmetry which would in general require constant but non-zero curvature. The question is what is the structure of \textit{free} state of spacetime, flat or constantly curved? It stands to reason that the basic structure should remain the same whether it is in \textit{free} state or matter filled state. Then for the former the curvature is homogeneous while for the latter it is inhomogeneous. This is the general dynamical characterization of spacetime. \\

Let us end with a speculation. Had Einstein followed the Bianchi identity, he couldn't have helped taking $\Lambda$ seriously as a new constant of spacetime and gravity? It is then conceivable that he could have made a profound prediction that the Universe may indeed experience accelerated expansion some time in the future. From the greatest blunder, it would have turned into the cause for the most remarkable cosmological prediction. When it would have been actually borne out by the observations \cite{accel}, we would have once again saluted his genius.  Alas it didn't happen this way! \\



\end{document}